\documentclass[prl,preprint,aps]{revtex4}

\usepackage{graphicx}
\usepackage{amsmath}

\begin{document}


\title{Chaotic Microcavity Laser with Low threshold and Unidirectional Output}

\author{Q. Song$^1$, H. Cao$^1$, B. Liu$^2$, S. T. Ho$^2$, W. Fang$^3$, and G. S. Solomon$^3$}
 
\address{$^1$ Department of Applied Physics, Yale University, New Haven, CT 06520-8482 \\
$^2$ Department of Electrical Engineering and Computing Science, Northwestern University, Evanston, IL, 60208 \\ 
$^3$ Atomic Physics Division, NIST, Gaithersburg, MD 20899-8423. } 

\begin{abstract}
Here we report lasing action in lima\c{c}on-shaped GaAs microdisks with quantum dots (QDs) embedded. Although the intracavity ray dynamics is predominantly chaotic, high-$Q$ modes are concentrated in the region $\chi > \chi_c$ as a result of wave localization. Strong optical confinement by total internal reflection leads to very low lasing threshold. Our measurements show that all the lasing modes have output in the same direction, regardless of their wavelengths and intracavity mode structures. This universal emission direction is determined by directed phase space flow of optical rays in the open chaotic cavity. The divergence angle of output beam is less than 40 degree. The unidirectionality proves to be robust against small deviations of the real cavity shape and size from the designed values. 
\end{abstract}

\maketitle

The microdisk laser is a promising candidate for light sources in integrated photonic circuits because of its low lasing threshold and in-plane output \cite{chang,vahala}. One major drawback of a circular microdisk laser is its isotropic output. Cavity shape has been employed as a design parameter in optimizing microlaser performance, e.g. deformation from circularity to achieve anisotropic output \cite{nockel,gmachl,schwefel}, which is an easy and cost-effective way compared to waveguide coupling which requires accurate positioning of a waveguide next to or underneath the cavity \cite{zhang,kippenberg,choi}.  However, 
deformed microdisk lasers produce output beams in multiple directions. For many applications, e.g., single-photon sources, unidirectional emission is essential. Only a few disk shapes such as spiral and rounded isosceles generate light output in a single direction \cite{chern,kurdo,bary}, but they suffer severe quality ($Q$) spoiling which leads to very high lasing threshold. 
Employing the characteristic modifications of spatial mode structures near avoided resonance crossings was also proposed to achieve both high Q and unidrectional emission \cite{wiersigPRA06}. Unfortunately it works only for particular modes and the existence of nearly degenerate modes with different far-field patterns smears out the output directionality. 

Utilizing a chaotic microcavity is another possible way \cite{wiersigPRL08}. An intuitive expectation for modes of a chaotic dielectric cavity is low $Q$ factor and pseudorandom emission pattern, because an optical ray undergoing chaotic diffusion in phase space can quickly reach the leaky region (where the incident angle $\chi$ at cavity boundary is less than the critical angle $\chi_c$ for total internal reflection) and escape from the cavity by refraction. However, wave localization along unstable periodic orbits (UPOs) creates scarred resonances, which can have high quality factors if the UPOs are located beyond the leaky region in phase space \cite{fang,lee04}. Emission directionalities of these high-$Q$ scarred resonances are determined by the unstable manifolds of UPOs, which guide optical rays out of the cavity in certain preferred directions \cite{schwefel,lee,shin}. Hence, an open chaotic microcavity can produce directional emission via phase space flow along the unstable manifolds \cite{shinohara}. Moreover, many high-$Q$ scarred resonances may have similar far-field patterns if their corresponding UPOs have closely-nested unstable manifolds \cite{schwefel,lee07}. The chaotic microcavity lasers that have been fabricated so far have output beams in multiple directions \cite{haraPRE,leb}. Wiersig and Hentschel recently showed in their calculations that a lima\c{c}on cavity can support high-$Q$ modes with unidirectional emission. Here we report lasing in lima\c{c}on-shaped GaAs microdisks with InAs quantum dots (QDs) embedded. Although the intracavity ray dynamics is predominantly chaotic, the high-$Q$ modes are concentrated in the region $\chi > \chi_c$ as a result of wave localization. Strong optical confinement by total internal reflection leads to very low lasing threshold, which is comparable to that of a circular disk of similar size. Our measurements show that all the lasing modes have output in the same direction, regardless of their wavelengths and intracavity mode structures. This universal emission direction is determined by directed phase space flow of optical rays in the open chaotic cavity. The divergence angle of output beam is less than 40 degree. The unidirectionality proves to be robust against small deviations of the real cavity shape and size from the designed values. 

The sample is grown on GaAs substrate by molecular beam epitaxy. The layer structure consists of 1000 nm Al$_{0.68}$Ga$_{0.32}$As and 265 nm GaAs. Inside the GaAs layer there are six monolayers of InAs QDs equally spaced by 25 nm GaAs barriers. Lima\c{c}on-shaped microdisks are fabricated with photolithography and two steps of wet etching. The first step is a nonselective etching with HBr, followed by a HF-based selective etching to undercut Al$_{0.68}$Ga$_{0.32}$As. Figure 1 shows the top-view and side-view scanning electron microscope (SEM) images of a microdisk. The disk shape is slightly deviated from the lima\c{c}on of Pascal, and it can be fitted as $\rho = R(1+ \epsilon \cos \theta) - d$, where $R$ = 2.28 $\mu$m, $\epsilon = 0.43$, and $d$ = 0.1 $\mu$m. The side-view SEM image illustrates smooth sidewall of GaAs disk. The 1000 nm long Al$_{0.68}$Ga$_{0.32}$As pedestal separates the GaAs disk from the substrate. To minimize its effect on lasing modes in the disk, the pedestal is etched to have a top lateral dimension of 620 nm .   

In the lasing experiment, the microdisk is cooled to 10K, and optically pumped by a mode-locked Ti:Sapphire laser at wavelength 800 nm. A long-working-distance objective lens focuses the pump beam to a single disk from the top. A small amount of emission from the microdisk is scattered by the disk boundary to the top. It is collected by the same objective lens and directed to a spectrometer. Figure 2(a) shows a spectrum of emission from the disk shown in Fig. 1 at the incident pump power of 18.1 $\mu$W. It features several narrow peaks. Figure 2(b) plots the intensity and linewidth of one peak at $\lambda = 923$ nm as a function of incident pump power $P_i$. When $P_i$ exceeds 2.5 $\mu$W, the peak intensity exhibits a rapid increase with pumping, meanwhile the linewidth drops quickly to 0.16 nm. This threshold behavior indicates the onset of lasing oscillation. The inhomogeneously broadened gain spectrum of InAs quantum dots allows lasing in multiple modes which are well separated in wavelength. The pump power required to reach the lasing threshold is comparable to that of a circular microdisk with the same area \cite {HuiAPL00}. The low lasing threshold suggests efficient confinement of light in the lima\c{c}on cavity.  
 
To measure the far-field pattern of laser emission from a lima\c{c}on cavity,  we fabricate a ring structure around each microdisk. The ring is centered at the disk and has a radius of 34  $\mu$m, which is much larger than the disk radius $R$. The in-plane laser emission from the disk propagates to the ring and is scattered out of the plane. The scattered light pattern is imaged by the objective lens to a CCD camera.  Figure 3(a) shows an image taken at $P_i$ = 18.1 $\mu$W. To obtain a clear image of weak scattered light along the ring, we use long integration time for the CCD camera, which causes an over-exposure of the disk itself. Since the ring radius exceeds $4R^2/ \lambda$, the scattered light intensity along the ring reflects the far-field emission pattern of the microdisk. It shows the laser output from the lima\c{c}on disk is mostly in one direction. The red line in figure 3(b) plots the scattered light intensity as a function of polar angle $\theta$.  The unidirectional emission is centered around  $\theta=0$ with a width of 40$^{\circ}$. 
The emission spectrum taken simultaneously with the image [Fig. 2(a)] reveals multi-mode lasing. Hence, the unidirectional emission shown in Fig. 3(a) comes from all lasing modes. To find out the far-field pattern of individual lasing modes, we place a narrow bandpass filter in front of the CCD camera to select only one mode. By tuning the filter's wavelength, we image different lasing modes. The far-field patterns of two lasing modes at 909 nm and 923 nm are also plotted in Fig. 3(b). They are similar to the far-field pattern of total laser emission except for a small variation in the angular distribution of output intensity. Our data confirm that all the lasing modes have  output beams in the same direction with similar divergence angle.  

To identify the lasing modes, we have performed numerical simulations of actual microdisks that are measured. Since its radius is much larger than its thickness, the microdisk can be regarded as a two-dimensional cavity with an effective index of refraction $n_{eff}$. For transverse electric (TE) polarization (the electric field parallel to the disk surface), we calculate $n_{eff}$ = 3.13. Experimentally the lasing modes are TE-polarized, and correspond to high-$Q$ modes of frequencies within the gain spectrum. After obtaining the exact shape of the microdisk from the SEM image, we compute the high-$Q$ TE resonances by solving the Maxwell's equations with the finite-difference time-domain (FDTD) method. Due to limited accuracy in determining the refractive index of GaAs at low temperature, the frequencies of high-$Q$ modes do not match exactly those of lasing modes.  Figures 4(a) and (b) plot the spatial profiles of two high-$Q$ modes in the microdisk shown in Fig. 1. The mode at $\lambda = 928$ nm has $Q$ = 94000, and the one at 945 nm has $Q$ = 15000. Figures 4 (c) and (d) are the Husimi phase-space projections of these two modes. It is evident from Fig. 4(d) that the mode at 945 nm is localized on an unstable periodic orbit with four bounces from the cavity boundary. The locations of the bounces are marked by green dots in Fig. 4(d). The orbit itself is drawn in the inset of Fig. 4(d). Hence, the mode at 945 nm is a scar mode. From its spatial profile and Husimi projection, the mode at 928 nm is more like a whispering-gallery (WG) mode than a scar mode. However, the whispering-gallery trajectories in a lima\c{c}on cavity with $\epsilon = 0.43$ are confined to a tiny region $|\sin \chi| \geq 0.99$ \cite{wiersigPRL08}, while the mode at 928 nm is concentrated mostly between $|\sin \chi| = 0.74$ and  $|\sin \chi| = 0.93$. Strictly speaking, the mode at 928 nm is not a WG mode. No stable islands exist below the WG region and above the leaky region. Hence, we believe the mode at 928 nm is formed by dynamical localization \cite{starykh,Wei}. Namely, the chaotic diffusion in phase space is suppressed by destructive interference, which leads to the formation of WG-like mode. As shown in Figs. 4 (c) and (d), both the scar mode at 945 nm and the WG-like mode at 928 nm are localized well above the leaky region ($\sin \chi < 1/n_{eff}$). Their exponentially small intensities in the leaky region  explain their high $Q$ factors. Another consequence of wave localization in the region $\chi > \chi_c$ is that the modes have very low field intensities in the disk center. Thus degradation of their $Q$ factors by the Al$_{0.68}$Ga$_{0.32}$As pedestal is negligible. We have calculated many modes and confirmed that all the high-$Q$ modes are concentrated well above the leaky region due to wave localization.   

Figures 4(e) and (f) plot the angular distributions of far-field intensities for the two high-$Q$ modes of $\lambda =$ 928 nm and 945 nm. Although their intracavity mode structures are different, the far-field patterns are similar. Their outputs are predominantly in a single direction $\theta$ = 0. Our numerical simulations verify that all high-$Q$ modes have outputs in the same direction.  To understand the universal directionality of high-$Q$ modes, we take a close inspection of their Husimi projections in the leaky region. In Figs. 4(c) and (d), the color-enhanced Husimi projections below the critical line $\sin \chi = 1 / n_{eff}$ reveal that the escape routes from the cavity are nearly identical for the scar mode and the WG-like mode. Wiersig and Hentschel showed in \cite{wiersigPRL08} that light escape for high-$Q$ scar modes in the lima\c{c}on cavity is along the unstable manifolds. The UPOs above the leaky region have closely-nested unstable manifolds, which cross the critical line at almost the same point. Hence, the output beams have identical direction. Our calculations show that the high-$Q$ modes are not necessarily scar modes, they can be WG-like modes formed by dynamical localization. Although they do not correspond to specific UPOs, the WG-like modes have similar unidirectional output as the scar modes. To confirm the output directionality is determined by chaotic ray dynamics, we have performed classical ray tracing in our microdisk whose shape is slightly deviated from the lima\c{c}on. The initial rays, with identical amplitudes, are uniformly distributed in the phase space above the critical line. At each reflection from the boundary, the amplitude of a ray is reduced according to the Fresnel law. Tracing of one ray is stopped after its amplitude falls below a certain value. The accumulated amplitude distribution in the leaky region, shown in Fig. 4(g), resembles the intensity distribution of the WG-like mode in Fig. 4(c) and the scar mode in Figs. 4 (d). The good agreement between the ray tracing result and the Husimi projection of high-$Q$ modes confirms that the universal directionality of output beam results from the classical ray dynamics.

In conclusion, we have achieved simultaneously low lasing threshold and unidirectional emission in a chaotic microcavity. The cavity shape is defined by the lima\c{c}on of Pascal, and the cavity size is smaller than 5 $\mu$m. We observe multi-mode lasing thanks to the broad gain spectrum of InAs QDs. Wave localization above the leaky region in phase space ensures strong optical confinement and low lasing threshold. All the lasing modes have output beams in the same direction with a divergence angle less than $40^{\circ}$.  The emission directionality is determined by chaotic ray dynamics in the open cavity, and is robust against small deviation of cavity shape from lima\c{c}on. Such tolerance allows fabrication with standard photolithography and wet chemical etching, that would facilitate parallel mass production.

\pagebreak
{\bf Figure Captions}

Fig. 1: Scanning electron microscope images of a GaAs microdisk on top of a Al$_{0.68}$Ga$_{0.32}$As pedestal. (a) Top view, (b) side view. The GaAs disk is 265 nm thick and contains six layers of InAs QDs. 

Fig. 2: 
(a) Emission spectrum taken at the incident pump power of 18.1 $\mu$W. It features several narrow peaks. (b) Intensity (black dot) and spectral width (red cross) of one peak at $\lambda$ = 923 nm as a function of incident pump power $P_i$. A threshold is clearly seen at $P_i$ = 2.5 $\mu$W where the slope of peak intensity (marked by black line) changes and the spectral width drops quickly.

Fig. 3: Measured directionality of laser emission from the microdisk shown in Fig. 1. (a) An image of laser emission from the microdisk scattered by the ring surrounding the disk. The ring radius is much larger than the disk radius. The incident pump power is 18.1 $\mu$W. The emission spectrum taken simultaneously with the image is shown in Fig. 2(a). (b) Far-field angular distributions of emission intensities of all lasing modes (red solid line), as well as of two lasing modes at $\lambda$ = 909 nm (blue dashed line) and 923 nm (black shot dash line), respectively. All lasing modes have output beams in the same direction ($\theta $ is the angle in limacon fitting formula) with a divergence angle of 40$^{\circ}$.

Fig. 4: Numerical simulation results of high-$Q$ modes in the microdisk of shape equal to that in Fig. 1. Calculated spatial intensity distributions (a, b), Husimi phase space projections (c, d), and far-field emission patterns (e, f) of two high-$Q$ modes at $\lambda$ = 928 nm and 945 nm, respectively. The horizontal axis of (c) and (d) represents the length along the cavity boundary from the point $\theta = 0$ normalized by the cavity perimeter $S$, and the vertical axis corresponds to $\sin \chi$, where $\chi$ is the incident angle at the cavity boundary. The spatial pattern and Husimi projection of the mode at 928 nm are similar to those of a whispering-gallery mode. The mode at 945 nm is localized on an unstable periodic orbit with four bounces from the cavity boundary. The bounces are marked by the green dots in (d) and the orbit itself is drawn in the inset of (d). In (c) and (d) the intensities of Husimi projections in the leaky region $\sin \chi < 1 / n_{eff}$ are enhanced to illustrate the escape route of light from the cavity. (g) is the distribution of optical rays in the leaky region obtained by classical ray tracing. It agrees well with the Husimi projections of high-$Q$ modes, illustrating the universal output directionality results from chaotic ray dynamics in the open cavity.

\pagebreak
\begin{figure}[htbp]
	\centering
		\includegraphics[width=0.3\textwidth]{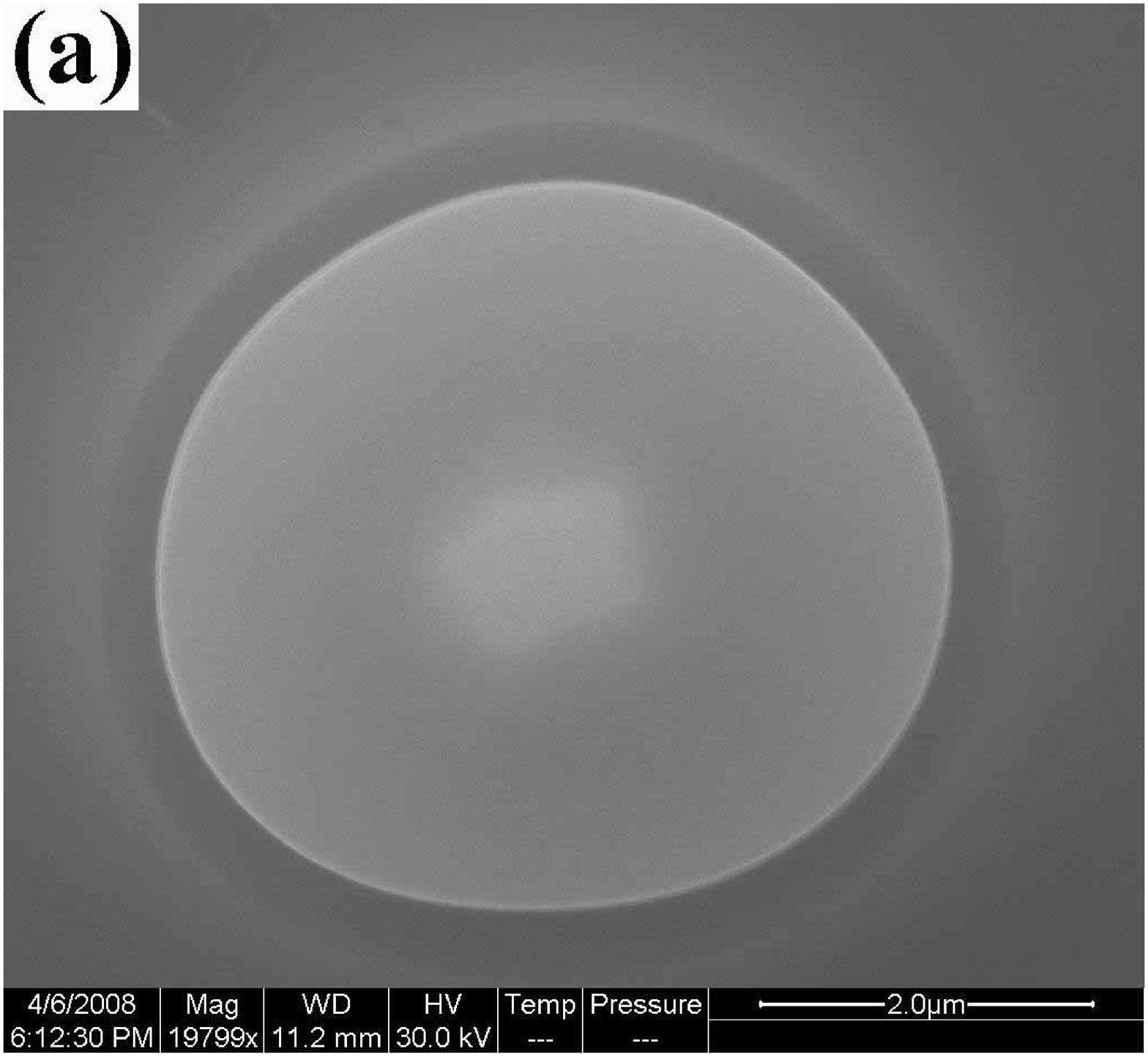}
		\includegraphics[width=0.3\textwidth]{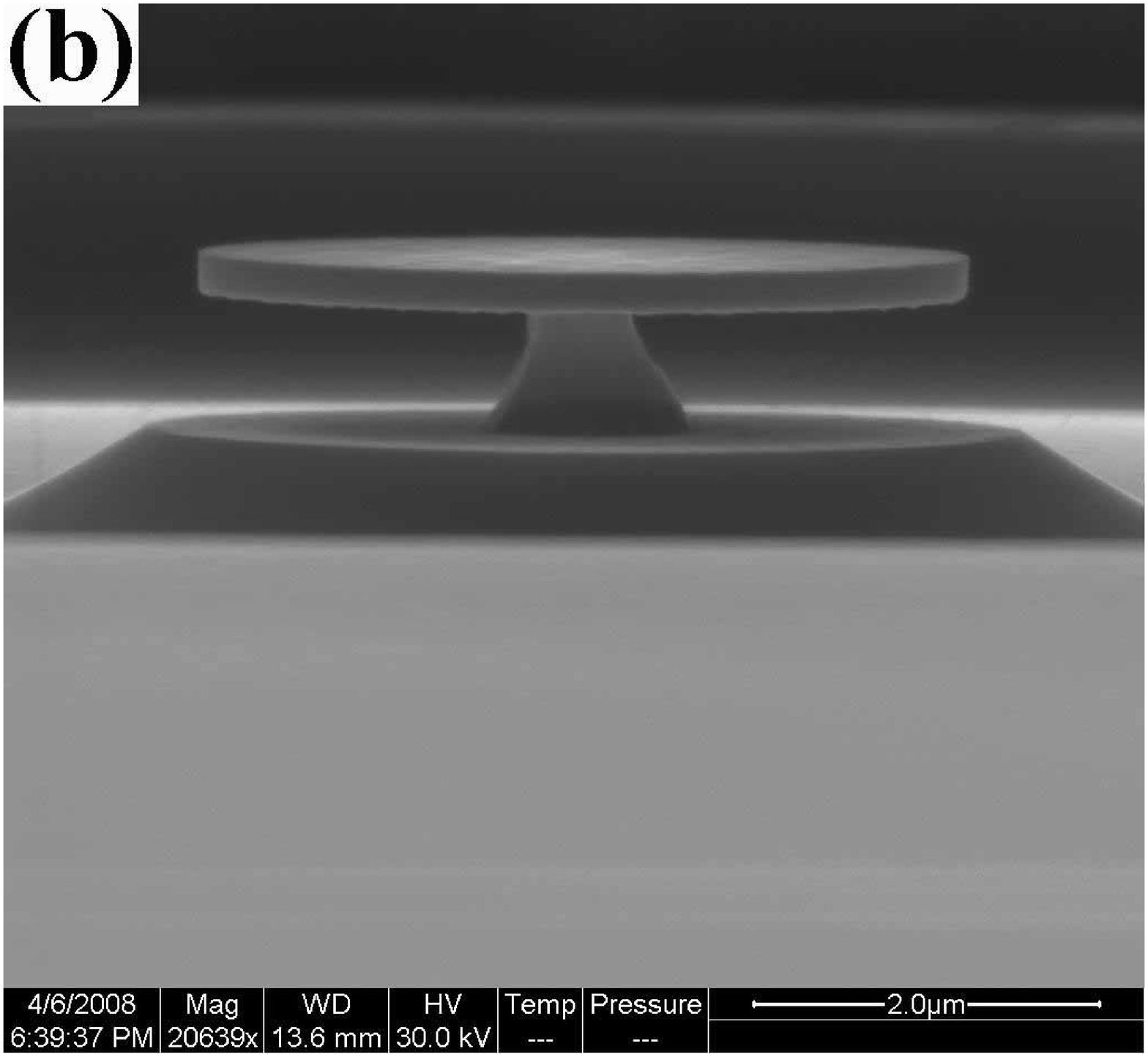}
\end{figure}
Fig.1 Song et al.
\pagebreak
\begin{figure}[htbp]
	\centering
		\includegraphics[width=0.4\textwidth]{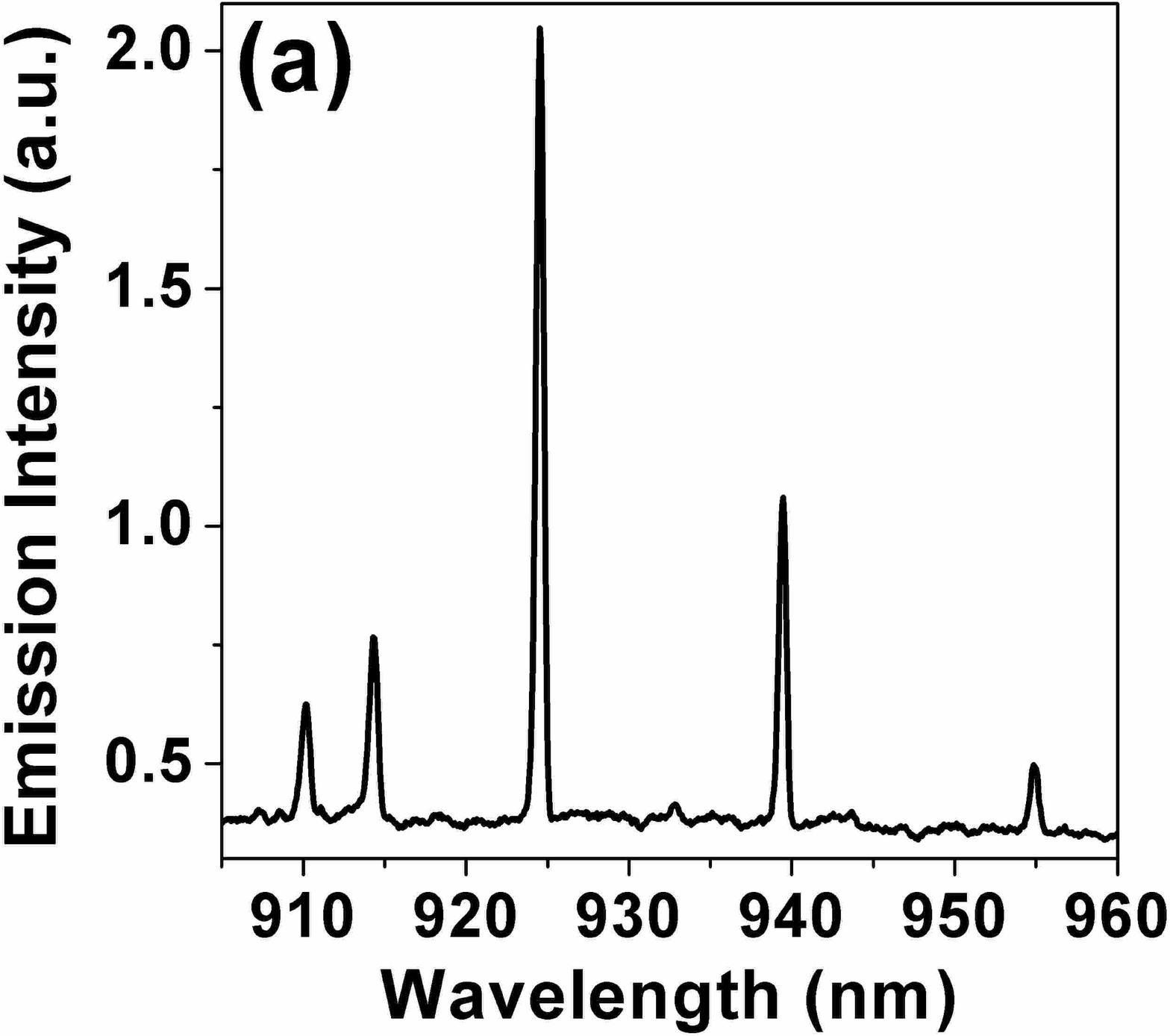}
		\includegraphics[width=0.4\textwidth]{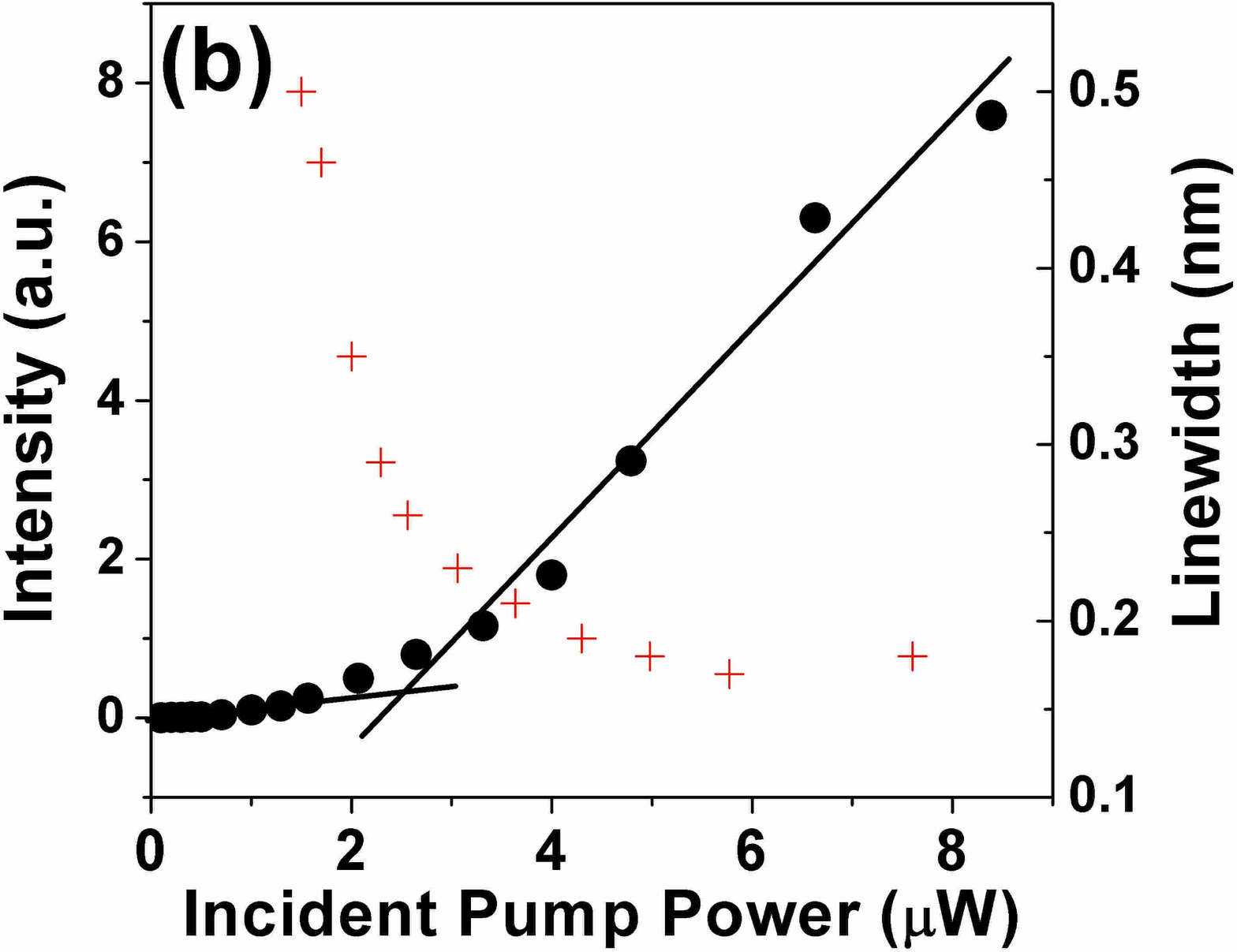}
\end{figure}

Fig.2 Song et al.

\pagebreak

\begin{figure}[htbp]
	\centering
		\includegraphics[width=0.27\textwidth]{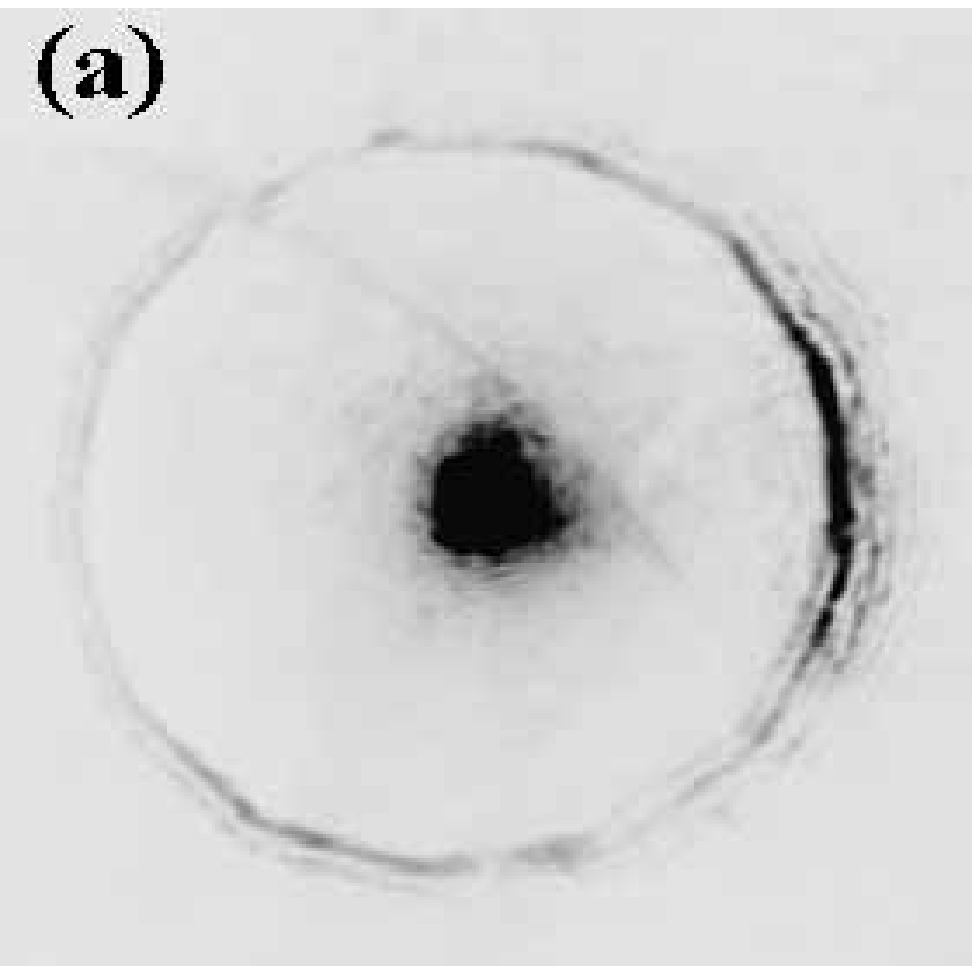}
		\includegraphics[width=0.27\textwidth]{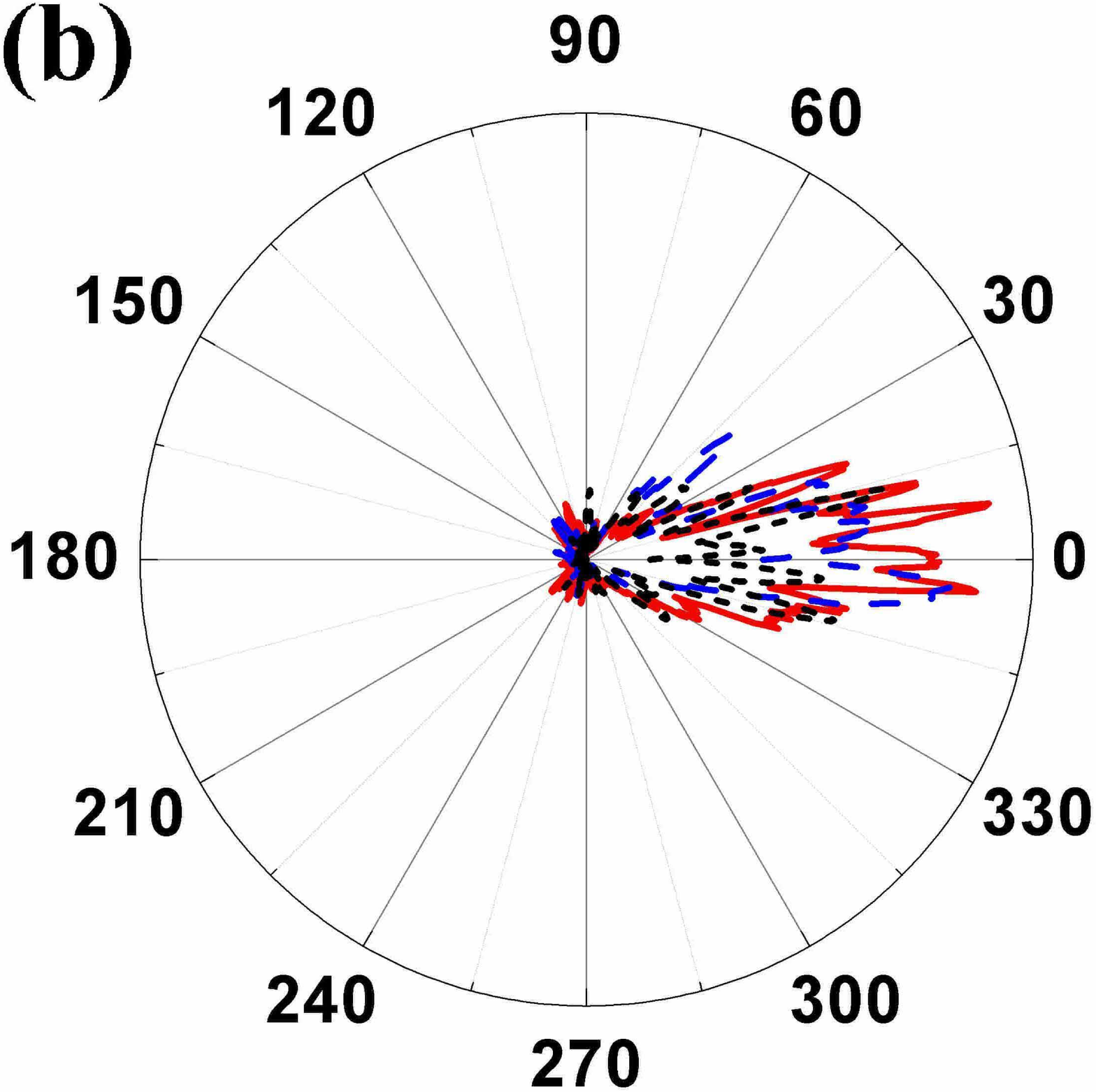}
\end{figure}

Fig.3 Song et al.

\pagebreak

\begin{figure}[htbp]
	\centering
	\begin{tabular}{c c c}
	 \begin{tabular}{c}
	 \includegraphics[width=0.195\textwidth]{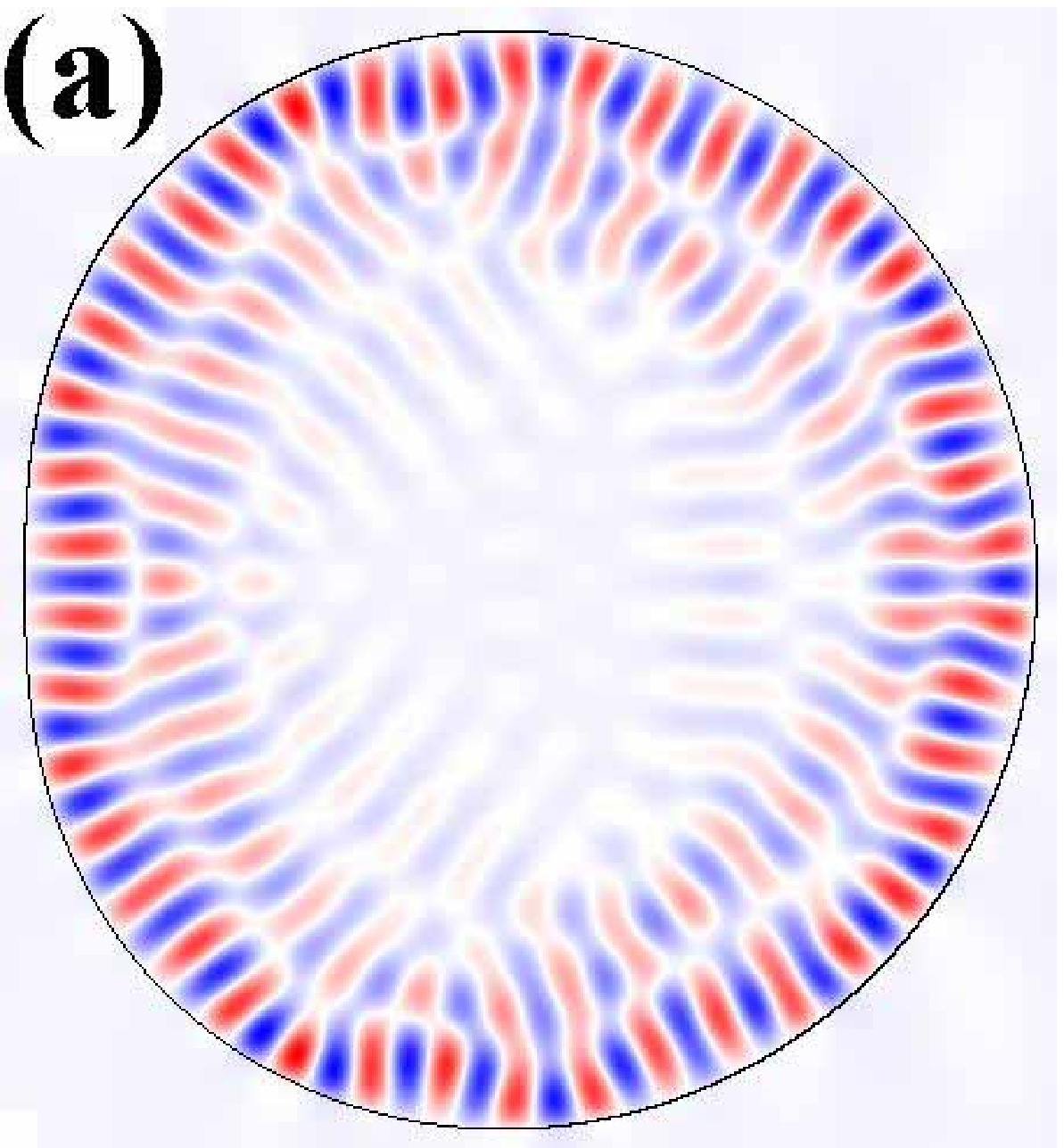} \\
	 \includegraphics[width=0.195\textwidth]{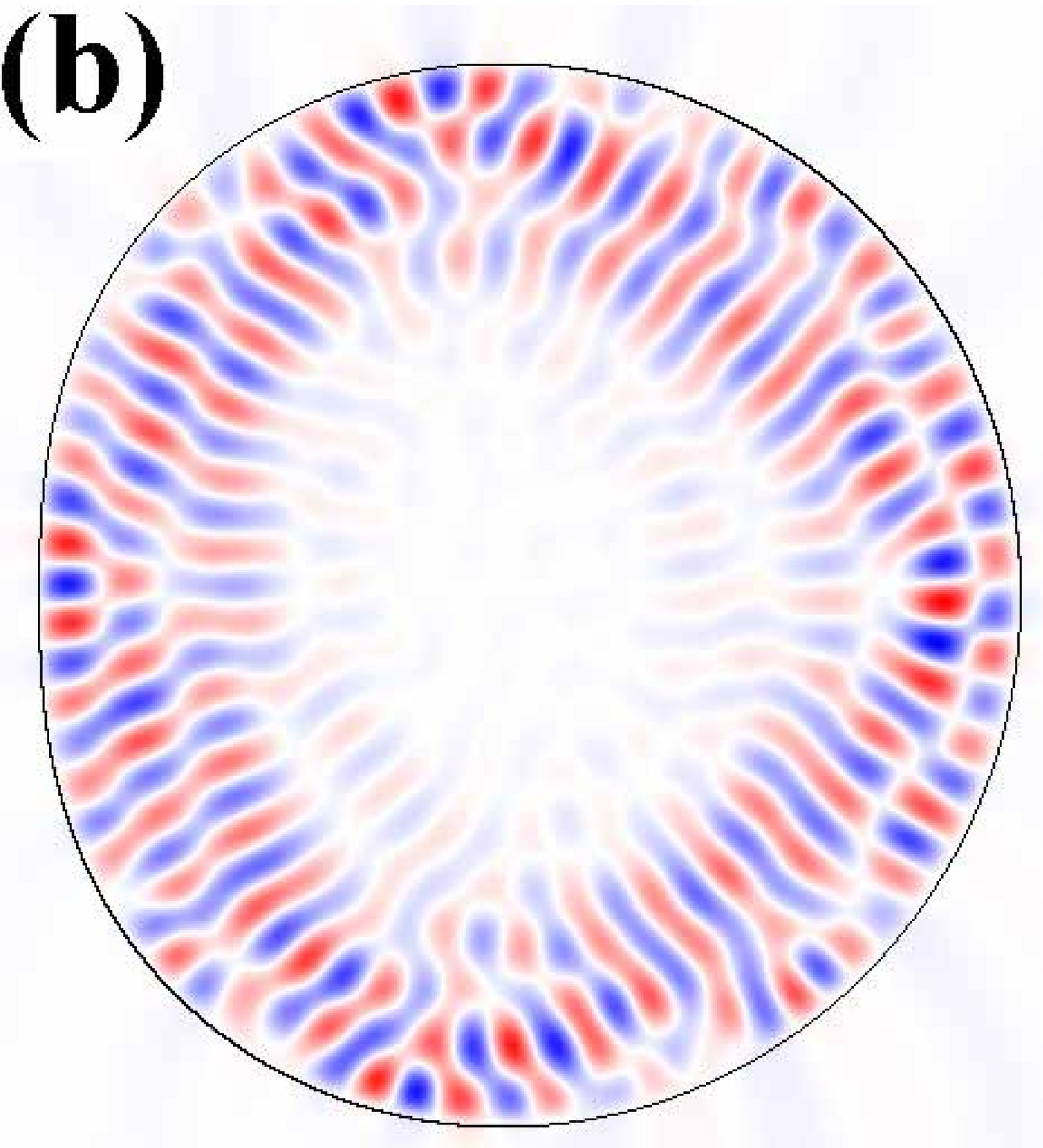}
	 \end{tabular} &
	 \begin{tabular}{c}
	 \includegraphics[width=0.315\textwidth]{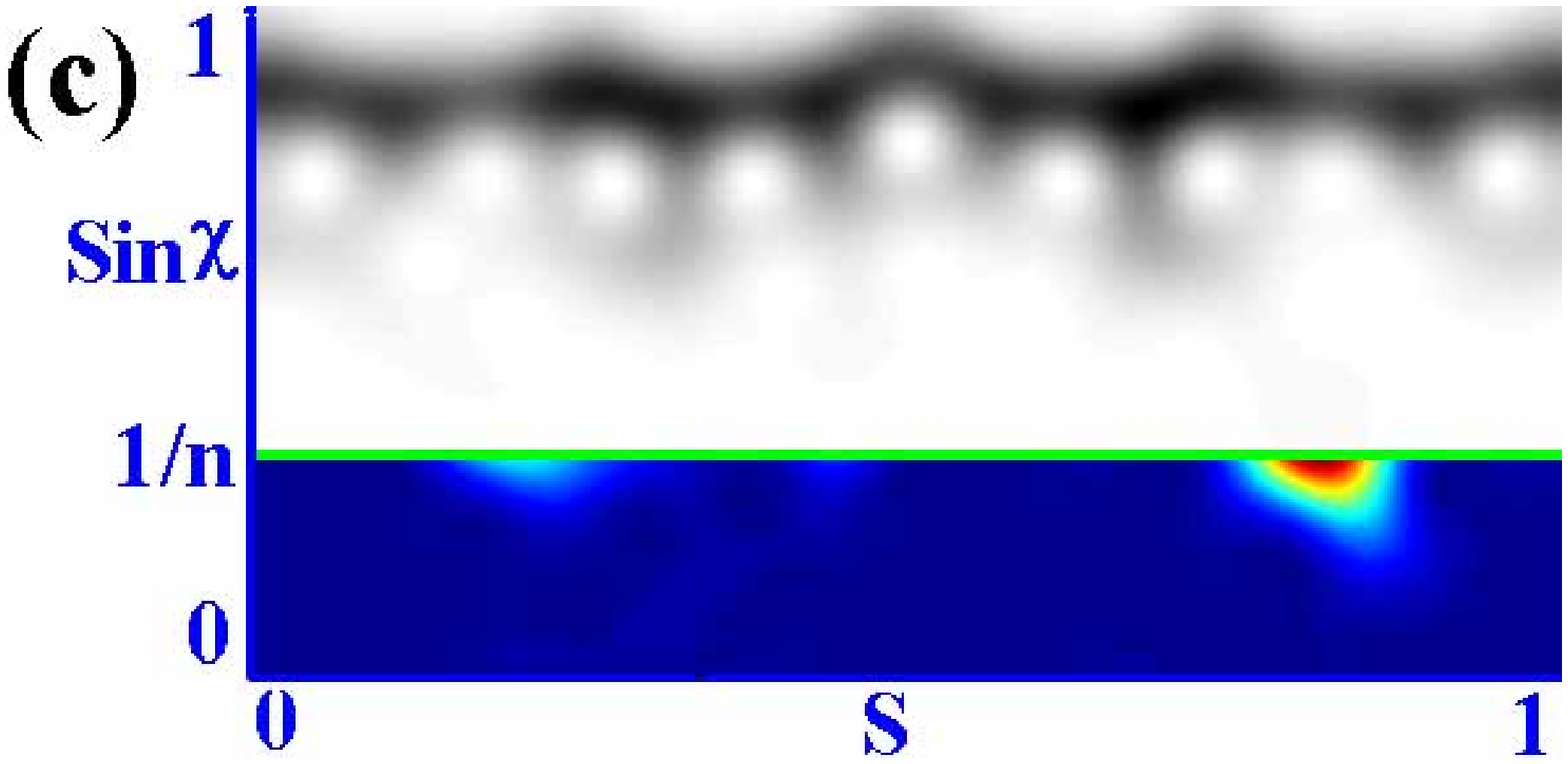} \\
	 \includegraphics[width=0.3\textwidth]{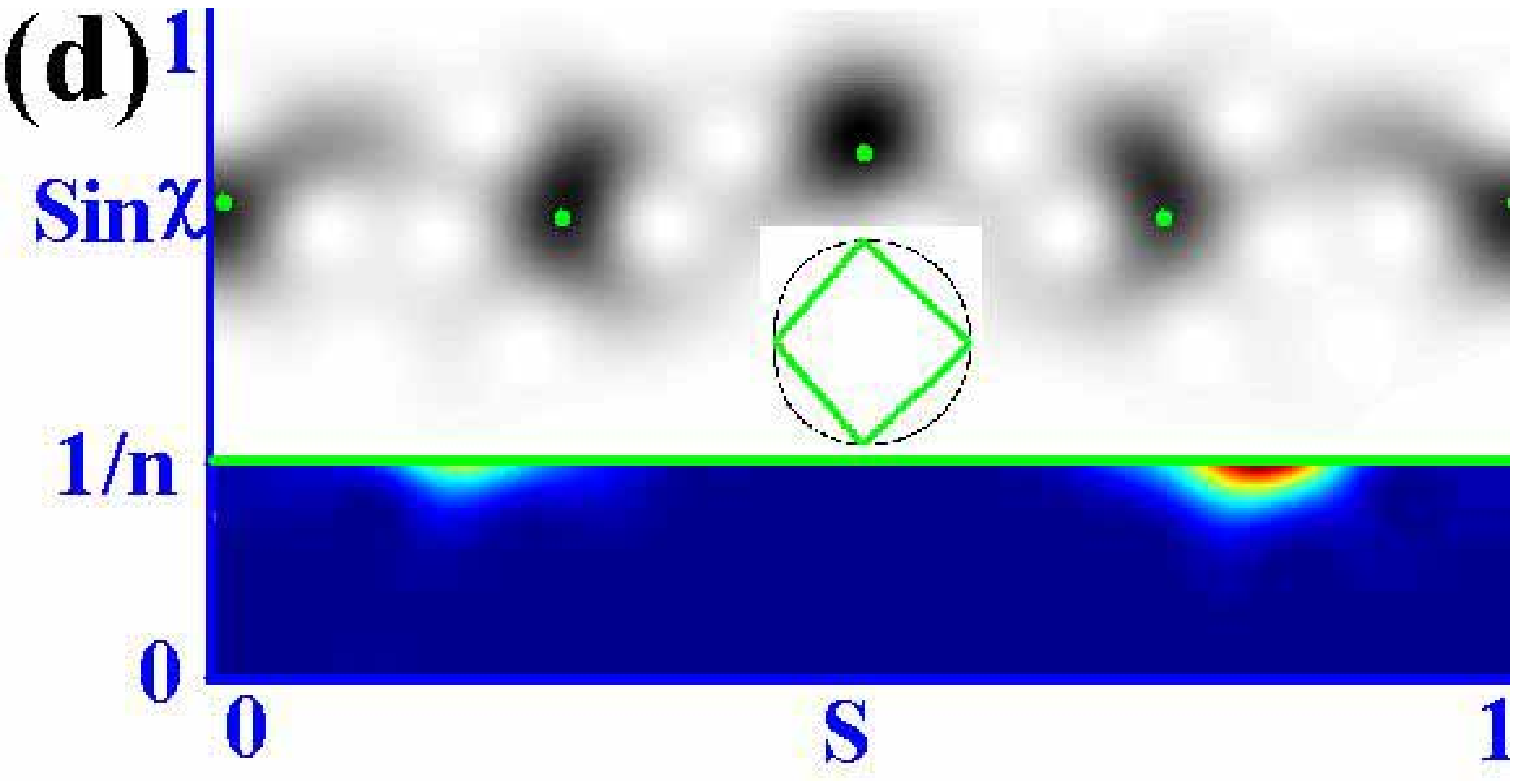} \\
	 \includegraphics[width=0.3\textwidth]{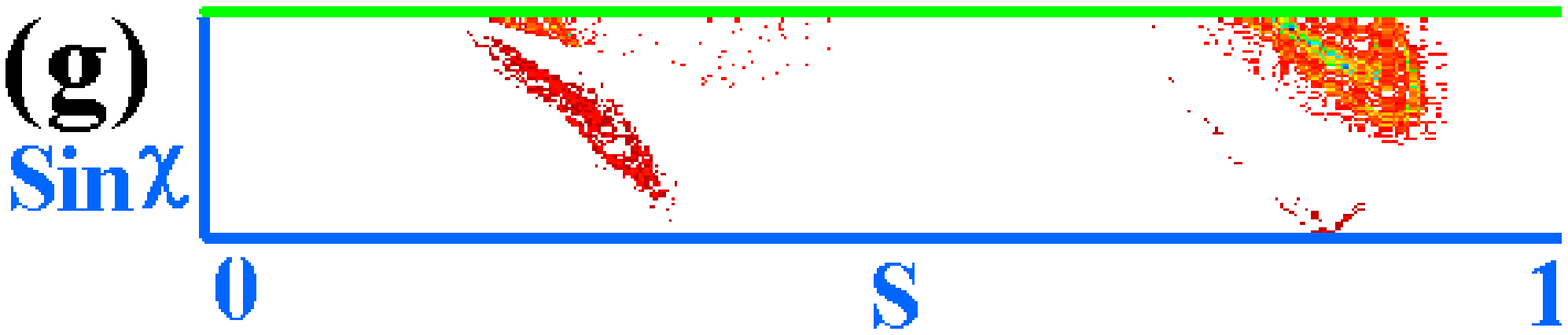}
	 \end{tabular} &
	 \begin{tabular}{c}
	 \includegraphics[width=0.2\textwidth]{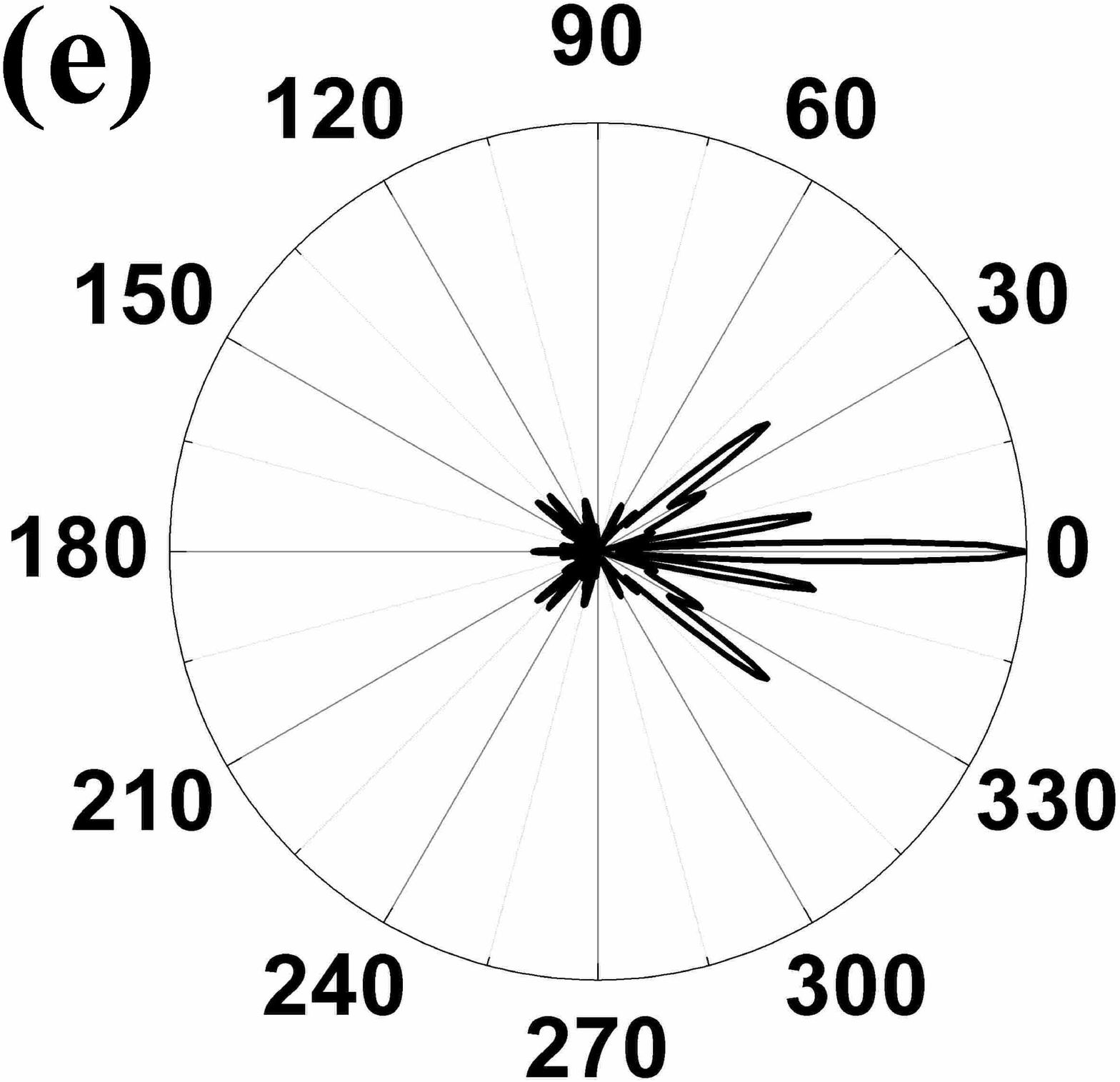} \\
	 \includegraphics[width=0.2\textwidth]{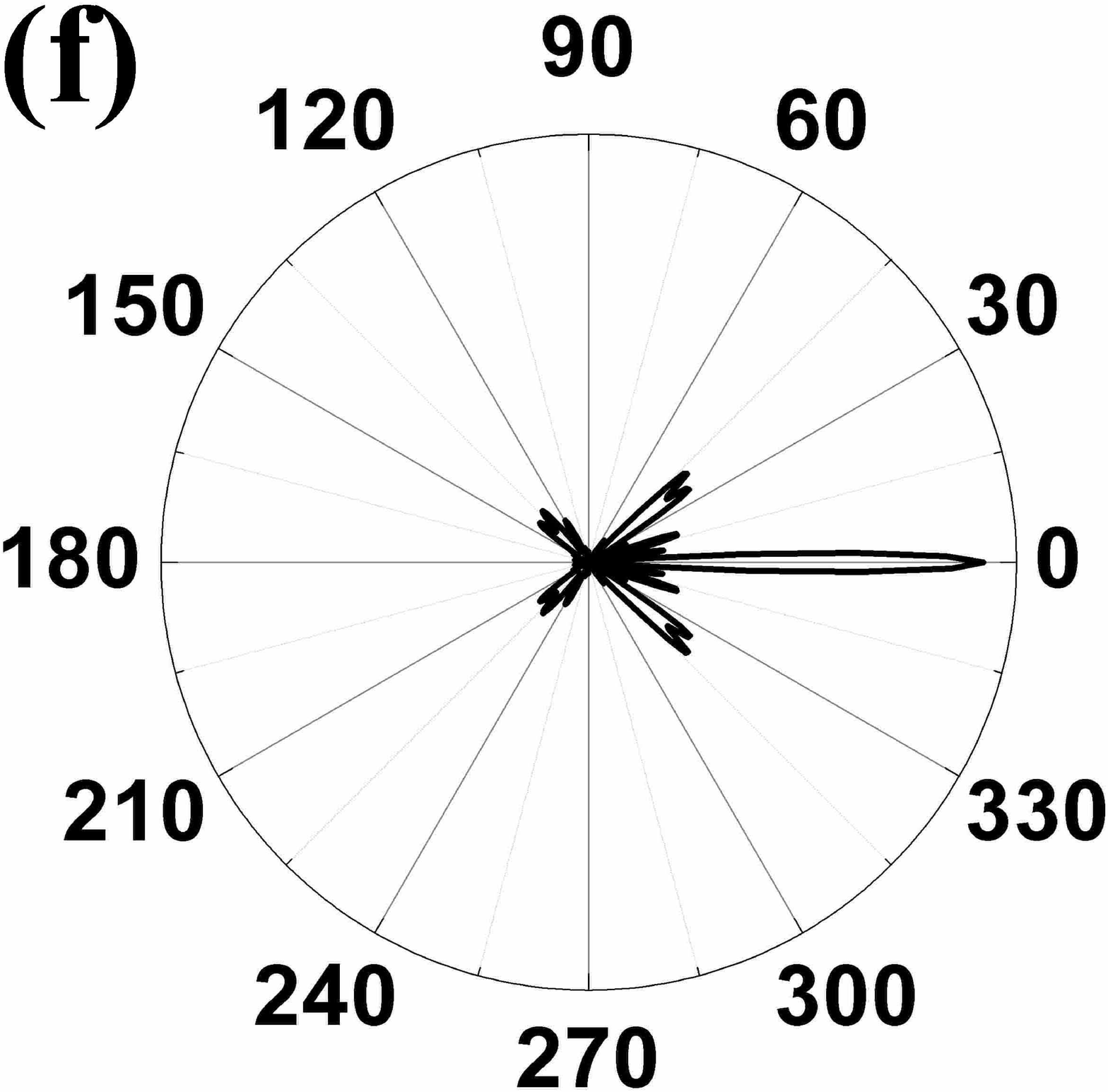}
	 \end{tabular} 
	\end{tabular}	
\end{figure}

Fig.4 Song et al.


\end{document}